\documentclass[12pt]{iopart}
\usepackage{graphicx}
\usepackage{epsfig} 

\begin{document}

\title[On O-X mode conversion in a cold magnetized 2D inhomogeneous plasma ...]
{On O-X mode conversion in a cold magnetized 2D inhomogeneous plasma in the electron cyclotron frequency range}

\author{A Yu Popov}

\address{Ioffe Physico-Technical Institute, St.Petersburg,
Russia}
\ead{a.popov@mail.ioffe.ru}
\begin{abstract}

In this paper a reduced set of the partial differential wave
equations valid in the conversion layer close to O-mode cutoff
surface and accounting for the magnetic field 2D inhomogeneity with
no restriction to an angle between the toroidal direction and the
magnetic field direction is derived. An integral representation of a
solution to the derived set of equation is given. For the particular
case of small angle between O mode cut-off surface and X mode
cut-off surface an explicit expressions for both the electric field
components and the conversion coefficients are obtained and its
properties are considered in details.

\end{abstract}

\maketitle

\section{Introduction}
Nowadays the electron Bernstein waves (EBWs) having no density
cut-offs and effectively damped even at high electron cyclotron
harmonics are considered as the most promising candidate to provide
an auxiliary heating and current drive in a dense plasma of a
spherical tokamaks and stellarators~\cite{Ref1, Ref2, Ref3}. The EBW
could be excited via direct conversion of X mode to Bernstein mode
in a vicinity of the upper hybrid resonance (UHR) or via so-called
O-X-B scheme. The efficiency of last scheme, as it was demonstrated
theoretically~\cite{Ref4, Ref5, Ref6} in the frame of 1D slab model,
is determined by the efficiency of O to X mode conversion, which can
reach 100 percent value at the certain parallel refractive index
being constant in slab geometry. Due to, in a real spherical
tokamak's configurations, where the poloidal inhomogeneity of the
magnetic field is important, both the parallel refractive index is
no longer constant and the components of the dielectric tensor are
functions of two co-ordinates, an analysis of the full-wave
equations in the frame of 2D model is important. The first attempt
to consider 2D model of OX conversion has been undertaken a couple
of years ago~\cite{Ref7}. The main conclusion provided by author of
~\cite{Ref7} concerning the absence of the O mode reflection from
the O-mode cut-off surface seems to be quite doubtful and lacks
support from the last two papers devoted to this topic ~\cite{Ref8,
Ref9}. Unfortunately, the OX mode conversion in ~\cite{Ref8} has
been considered in the frame of the oversimplified model ignoring as
it does both the poloidal magnetic field and, as result, varying of
the parallel refractive index on the magnetic surface. These effects
were taken into account in ~\cite{Ref9}, where the the explicit
expressions for the OX and XO conversion coefficients have been
obtained simultaneously with~\cite{Ref8} but for
$\varphi=\arctan{B_\theta/B_z}\ll\pi$ ($B_\theta$ and $B_z$ are the
poloidal and toroidal components of the magnetic field and $\varphi$
is an angle between the toroidal direction and the magnetic field).

Unlike the assumption used in ~\cite{Ref9}, $\varphi\ll\pi$ , in
spherical tokamak's plasma $\varphi$ is no longer small as it is
demonstrated in figure 1 for a typical MAST tokamak discharge.
Therefore, straightforward using the results of~\cite{Ref9}, without
mentioning paper ~\cite{Ref8}, to describe OX conversion in
spherical tokamaks seems to be overhasty. Because of the intensive
efforts are paid to provide an auxiliary heating in spherical
tokamaks using OXB scheme, calculation of the OX conversion
coefficient in the realistic 2D model is of great interest. In this
paper a reduced set of the partial differential wave equations valid
in a vicinity of the O-mode cutoff and accounting for the magnetic
field 2D inhomogeneity with no restriction to an angle between the
toroidal direction and the magnetic field are derived. A solution to
the reduced set of equations has been obtained and it properties are
considered in details.

\section{Physical model}
There are three effects that remain beyond the scope of the present
paper. We neglect, first, the curvature of the magnetic field line
at the magnetic surface because of its local radius $R_f$ is
considerably greater than the beam radius $\rho$, second, the
curvature of the magnetic flux surfaces assuming high localization
of the conversion region, third, the magnetic field shear which is
not important for OX conversion~\cite{Ref10}. We restrict ourselves
to the case of not extremely strong plasma density inhomogeneity
$L_n$ when geometrical optics can be applied except possibly near
cut-off layer or possibly mode conversion layer. One introduces two
Cartesian co-ordinate systems $\left(x,y,z\right)$ and
$\left(x,\zeta,\xi\right)$ with distances scaled in the units
$c/\omega$ and their origin located at the O-mode cut-off surface
(figure 2). The co-ordinates $x$, $y$ and $z$ imitate the flux
surface label, the poloidal and the toroidal co-ordinates,
respectively, and axes ${\bf e}_\zeta$ and ${\bf e}_\xi$ are along
the magnetic field and perpendicular to it on the O-mode cut-off
surface, respectively (figure 3). The transformation from $y$, $z$
components to $\zeta$, $\xi$ components convenient to represent the
Maxwellian equations is
\begin{eqnarray}
y=\cos(\varphi)\zeta+\sin(\varphi)\xi\nonumber\\
z=-\sin(\varphi)\zeta+\cos(\varphi)\xi.
\end{eqnarray}
One introduces the set of the wave equations for a monochromatic
wave $\sim\exp\left(i\omega t\right)$ as
\begin{eqnarray}
{\bf\nabla}\times{\bf\nabla}\times {\bf E}={\bf \epsilon}\,\,{\bf
E},\,\,\, {\bf\epsilon}= \left(
\begin{array}{c}
\epsilon_+\,\,\,\,\,0\,\,\,\,\,0\\
0\,\,\,\,\,\epsilon_-\,\,\,\,\,0\\
0\,\,\,\,\,0\,\,\,\,\,\eta\\
\end{array}\right),\,\,
{\bf E}= \left(
\begin{array}{c}
E_+\\
E_-\\
E_\xi\\
\end{array}\right),
\end{eqnarray}
where $E_\pm=(E_x\pm iE_\zeta)/\sqrt{2}$,
$\epsilon_\pm=1-v/\left(1\pm q\right),$ $\eta=1-v,$
$v=\omega_{pe}^2/\omega^2,$ $q=|\omega_{ce}|/\omega$, $\omega_{pe}$
and $\omega_{ce}$ are electron Langmuir and cyclotron frequencies,
respectively.

Since a domain in a vicinity of O-mode cut off surface beyond the
tokamak mid-plane is of interest, one can expand the plasma
parameters, namely density $n(x)$ and magnetic field modulus
$B(x,y)$,  into the Taylor serious at $r_0$
\begin{eqnarray}
n(x)\simeq n_0\left(1+x/L_n\right)\nonumber\\
B(x,y)\simeq B_0(1+x/L_{bx}+y/L_{by}),
\end{eqnarray}
where
\begin{eqnarray}
1/L_n,\rho/L_{by}\ll 1,\\ L_{n}^{-1}=\partial \ln{n_e}/\partial
x|_{r_0}, L_{by}^{-1}=\partial \ln{B}/\partial y|_{r_0},\nonumber
\end{eqnarray}
are parameters being the first order quantity ($O(1)$) and
\begin{eqnarray}
1/L_{bx}\ll 1,\\ L_{bx}^{-1}=\partial \ln{B}/\partial
x|_{r_0}\nonumber
\end{eqnarray}
is one being the second order quantity ($O(2)$). In order to study
the properties of the waves in the mode conversion region we keep in
mind that the component $E_-$ is small compared to two others
$E_-\sim \left(E_+,E_\xi\right)/L_n$ there. Omitting in (2) the
terms being higher order quantity than the first one obtains
\begin{eqnarray}
\left(\frac{\partial^2}{\partial
\xi^2}+\frac{q_0}{1+q_0}+\frac{x}{L_n}-\frac{y}{L_{by}}\right)E_+ +
\frac{1}{\sqrt{2}}\left(\frac{\partial}{\partial x}+
i\frac{\partial}{\partial
\zeta}\right)\frac{\partial}{\partial\xi}E_\xi =0\\
\frac{1}{\sqrt{2}}\left(\frac{\partial}{\partial x}-
i\frac{\partial}{\partial \zeta}\right)\frac{\partial}{\partial
\xi}E_+  -\frac{x}{L_n} E_\xi=0,\nonumber
\end{eqnarray}
Due to tokamak symmetry we may assume that the wave fields vary as
$\exp{(in_zz)}$, where $n_z$ being constant is large. We would like
to construct a solution ${\bf E}=\left(E_+,E_\xi\right)$ of the
system (6). To this end we look to develop the integral
representation for required solutions of Laplace integral type
\begin{eqnarray}
{\bf E}\left(x,y,z\right)=\int_{C}\frac{dn_y}{2\pi}
\int_{-\infty}^\infty\frac{dn_z}{2\pi}\exp{(in_yy+in_zz)}{\bf
E}\left(x,n_y,n_z\right),
\end{eqnarray}
where ${\bf E}\left(x,n_y,n_z\right)$ is assumed analytic in some
domain and the path of integration in the $n_y$ plane is such that
the integrand vanishes rapidly at the ends of the contour $C$ or at
infinity. Next

\begin{eqnarray}
\left(-n_\xi^2+n^{opt
2}+\frac{x}{L_n}-\frac{i}{L_{by}}\frac{\partial}{\partial
n_y}\right)E_+ +
\frac{in_\xi}{\sqrt{2}}\left(\frac{\partial}{\partial x}-n_\zeta\right)E_\xi =0\\
\frac{in_\xi}{\sqrt{2}}\left(\frac{\partial}{\partial
x}+n_\zeta\right)E_+  -\frac{x}{L_n} E_\xi=0,\nonumber
\end{eqnarray}
where $n^{opt}=\sqrt{q_0/(1+q_0)}$ and
\begin{eqnarray}
n_\zeta=\cos(\varphi)n_y-\sin(\varphi)n_z\\
n_\xi=\sin(\varphi)n_y+\cos(\varphi)n_z.\nonumber
\end{eqnarray}
We cannot easily deal with the system (8) as it stands after
transformation, but recalling the conclusions of 1D theory
~\cite{Ref4, Ref5} that the effective conversion is possible if both
$n_\zeta\ll 0$ $O(1)$ and $n_\xi\sim \pm n^{opt}$ $O(0)$ we may
reduce it to a simpler form. Expanding $n_\xi$ around $n^{opt}$ and
$n_\zeta$ around $0$ we have found that
$n_y^0=\sin{(\varphi)}n^{opt}$ and $n_z^0=\cos{(\varphi)n^{opt}}$.
Keeping in (8) terms being the first order quantity one has
\begin{eqnarray}
\left(-2n^{opt}\delta
n_\xi+\frac{x}{L_n}-\frac{i}{L_{by}}\frac{\partial}{\partial \delta
n_y}\right)E_+ +
\frac{in^{opt}}{\sqrt{2}}\left(\frac{\partial}{\partial x}-\delta n_\zeta\right)E_\xi =0\\
\frac{in^{opt}}{\sqrt{2}}\left(\frac{\partial}{\partial x}+\delta
n_\zeta\right)E_+  -\frac{x}{L_n} E_\xi=0,\nonumber
\end{eqnarray}
where the explicit expressions for $\delta n_\xi$ and $\delta
n_\zeta$ could be found by varying (9). Making the backward
transformation with respect to $\delta n_y$ yields
\begin{eqnarray}
{\bf E}({\bf r})=\exp{\left(in_z^0z+i
n_y^{0}y\right)}\int_{-\infty}^\infty \frac{d\delta
n_z}{2\pi}\exp{\left(i\delta n_zz\right)}\times\nonumber\\
\int_{-\infty}^\infty
dy'{G\left(y-y'\right)\exp{\left(-i\tan{(\varphi)} \delta
n_zy'\right)}\tilde{{\bf
E}}(x,y',\delta n_z)},\,{\bf r}=(x,y,z)\\
G\left(y-y'\right)=\frac{\exp{\left(-i(y-y')^2
/R^2\right)}}{\sqrt{i\pi}R},\,\,R^2=4L_{by}n^{opt}|\sin(\varphi)|
\end{eqnarray}
where the electric field's components $\tilde{{\bf
E}}=\left(\tilde{E}_+,\tilde{E}_\xi\right)$ obey the system of
equations

\begin{eqnarray}
\left(\frac{x}{L_n}-\frac{y}{L_{by}}-2n^{opt}\cos{(\varphi)\delta
n_z}\right)\tilde{E}_+ +\frac{i}{\sqrt{2}}
\left(\frac{\partial}{\partial
x}+i\cos\varphi\frac{\partial}{\partial y}\right)n^{opt}\tilde{E}_\xi =0\nonumber\\
\frac{i}{\sqrt{2}}\left(\frac{\partial}{\partial
x}-i\cos\varphi\frac{\partial}{\partial y}\right)n^{opt}\tilde{E}_+
-\frac{x}{L_n}\tilde{E}_\xi=0.
\end{eqnarray}

Introducing new notations
$$\frac{2^{1/4}}{L_n^{1/2}q_0^{1/4}}\cdot
x,\left(\frac{y}{\cos(\varphi)}+2n^{opt}L_{by}\delta
n_z\right)\rightarrow x,y,\,\,\partial_\pm=\frac{\partial}{\partial
x}\pm i\frac{\partial}{\partial y}$$
$$
a=\frac{L_n}{L_{by}}\frac{
q_0\cos{(\varphi)}}{1+q_0},\,\,\tilde{E}_+,\tilde{E}_\xi\rightarrow
E_+,E_\xi,\,\,F=-i\frac{E_+}{\sqrt{1+q_0}}$$ we read (13) in the
compact form
\begin{eqnarray}
\label{eq2} \partial_+E_\xi &+& \left(x-ay\right)F=0\nonumber\\
\partial_-F &-& x E_\xi=0
\end{eqnarray}

In order to make the first step in analysis of the electric field
components behavior in the conversion region, we consider in the
next Section WKB solution to the system (14), namely we focus on the
quality analysis of the ray trajectory along which the energy of the
incident beam of the ordinary (or extraordinary) waves is carried
over. Although in the mode conversion region the WKB approximation
breaks down and a local full wave equations (14) to be solved, the
ray trajectory analysis could be useful.

\section{Ray Hamiltonian dynamics}
The ray representation of the wave field in four dimensional ray
phase space ${\bf r}=(x,y),$ ${\bf n}=-i\partial/\partial{\bf r}$ is
governed by Hamilton's equations:
\begin{eqnarray}
\frac{d{\bf r}}{ds}=-\frac{\partial D}{\partial {\bf n}}
\left|\frac{\partial D}{\partial {\bf
n}}\right|^{-1},\,\,\frac{d{\bf n}}{ds}=\frac{\partial D}{\partial
{\bf r}}\left|\frac{\partial D}{\partial {\bf n}}\right|^{-1},
\end{eqnarray}
where
\begin{eqnarray}
D\left({\bf r},{\bf n}\right)=\left(n_x^2+\delta
n_y^2-x(x-ay)\right)/2=0,
\end{eqnarray}
is the local dispersion relation and $s$ denotes the orbit
parameter. Since the general picture of the ray behavior in the
two-dimensional subspace $xy$ is of interest, let us re-parameterize
$s\rightarrow\tau$, where $d\tau = ds \left|\partial D/\partial {\bf
n}\right|^{-1}$. Substituting (16) into (15) we obtain
\begin{eqnarray}
\frac{d^2{\bf r}}{d\tau^2}=-\frac{\partial U\left({\bf
r}\right)}{\partial {\bf r}},
\end{eqnarray}
where $U=x(x-ay)/2$. The equations constituting (17) are coupled
with $a$ being the coupled coefficient. One can introduce the normal
co-ordinates ${\bf r}=\left(x,y\right)\rightarrow \left(u,v\right)$
reducing the system of equations (17) to two independent ones. The
set of linear independent solutions for either of the two is
\begin{eqnarray}
u\sim\exp{\left(\pm\nu\tau\right)},\,
\nu=\left(\sqrt{1+a^2}+1\right)^{1/2}/\sqrt{2},\nonumber\\
v\sim\exp{\left(\pm
i\omega\tau\right)},\,\omega=\left(\sqrt{1+a^2}-1\right)^{1/2}/\sqrt{2}.
\end{eqnarray}
The expressions (18), which show the projections of Hamiltonian rays
on the ($u,v$) plane, deserves a few comments. First, note the
oscillatory behavior along the $v$ direction, while the $u$-motion
displays the influence of a retarding force. Second, due to
$|\nu|\gg|\omega|$ oscillations along the $v$ direction are not to
be clear-cut. Third, we can anticipate the full wave solution to the
set of equations (14) could be represented as the superposition of
the eigenmodes intrinsic to the confining potential ($\sim v^2$) in
the $v$ direction leading to oscillatory $v$ motion and the
superposition of the linear independent parabolic cylinder functions
along $u$ direction, which proper for the effective potential $\sim
-u^2$.  The full-wave solution to the set of equations (14) is
studied in the next Section.

\section{An integral representation of solution}

We would like to obtain the appropriate particular solution of the
system (14) matching to the WKB solution outside the mode conversion
region
\begin{eqnarray}
{\bf E}|_{u\rightarrow\pm\infty}\sim{\bf e}\exp{\left(\pm
iu^2/2\right)}\sum_{p=0}^\infty\psi_p(v),
\end{eqnarray}
where {\bf e} is the polarization vector, $\psi_p(v)$ is a set of
eigen functions to be found, $\exp{(-iu^2/2)}$ and $\exp{(iu^2/2)}$
correspond to incident O mode (OX conversion) and incident X mode
(XO conversion).

With that end in view we introduce two possible functional
substitutions~\cite{Ref9} providing the identity of the equations
constituting (14)
\begin{eqnarray}
F=
\left(x\pm\frac{i}{\sqrt{1-ia}}\partial_+\right)W^\pm\\
E_\xi=\left(\mp \frac{i(x-ay)}{\sqrt{1-ia}}+\partial_-\right)W^\pm
\end{eqnarray}
with $W^+$ and $W^-$ being required functions that satisfy the
equations
\begin{eqnarray}
\frac{\partial^2W^\pm}{\partial x^2
}+\frac{\partial^2W^\pm}{\partial y^2}+\left(x^2-axy\pm
i/\sqrt{1-ia}\right)W^\pm=0.
\end{eqnarray}
Let us use the transformation
\begin{eqnarray}
x=\left(1-N^2\right)^{1/4}\left(\cos{(\psi)}u-\sin{(\psi)}v\right),\nonumber\\
y=\left(1-N^2\right)^{1/4}\left(\sin{(\psi)}u+\cos{(\psi)}v\right),
\end{eqnarray}
where $\cos{(\psi)}=\left(1+N^2\right)^{-1/2},$
$\sin{(\psi)}=N\left(1+N^2\right)^{ -1/2}$ and
$N=-(\sqrt{1+a^2}-1)/a$, that makes the potential in the brackets in
(22) symmetrical
\begin{eqnarray}
\frac{\partial^2W^\pm}{\partial u^2}+\frac{\partial^2W^\pm}{\partial
v^2 }+\left(u^2-N^2v^2\pm \left(N+i\right)\right)W^\pm=0.
\end{eqnarray}
If $a<,>0$ (see figure 4) then $N>,<0$. Seeking the solutions to
(24) in the form
\begin{equation}
W^\pm(u,v)=\sum_{p=0}^\infty{W^\pm_p\left(u\right)\phi_p\left(\sqrt{|N|}v\right)},
\end{equation}
where the definition for $\phi_p(v)$ is given by
\begin{eqnarray}
\phi_p(\sqrt{|N|}v)=\frac{|N|^{1/4}}{\left(2^pp!\right)^{1/2}\pi^{1/4}}\exp{\left(-N^2v^2/2\right)}H_p\left(\sqrt{|N|}v\right),\\
\int_{-\infty}^\infty{\phi_p\left(\sqrt{|N|}v\right)\phi_k\left(\sqrt{|N|}v\right)dv}=\delta_{pk}\nonumber
\end{eqnarray}
and $H_p$ are Hermitian polynomials, we obtain the equation for
$W_p^+$ and $W_p^-$
\begin{eqnarray}
\frac{\partial^2W^\pm_p}{\partial
u^2}+\left(u^2-\frac{2\gamma^{\pm}_p}{\pi}\pm i\right)W^\pm_p=0,
\end{eqnarray}
where
\begin{eqnarray}
\gamma^{\pm}_p=\frac{\pi
|N|}{2}\left(2p+1\pm\frac{N}{|N|}\right).\nonumber
\end{eqnarray}
Since the properties of $\gamma^{\pm}_p$ are important for
subsequent analysis we list theirs below expressing its argument $N$
explicitly
\begin{eqnarray}
\gamma^{+}_p\left(N>0\right)=\gamma^{-}_p\left(N<0\right)=\pi
|N|\left(p+1\right),\nonumber\\
\gamma^{-}_p\left(N>0\right)=\gamma^{+}_p\left(N<0\right)=\pi
|N|p,\\
\gamma^{+}_{p-1}\left(N>0\right)=\gamma^{-}_{p}\left(N>0\right),\,\gamma^{+}_{p+1}\left(N<0\right)=\gamma^{-}_{p}\left(N<0\right).\nonumber
\end{eqnarray}

The solutions $W_p^+$ and $W_p^-$ to (27) matching to the WKB
solutions (19) for the incident ordinary and extraordinary waves
(see Appendix A) are
\begin{eqnarray}
W_{p}^+(u)=B_{p}^+D_{i\gamma^+_p/\pi}\left(\sqrt{2}\exp{\left(i\pi/4\right)}u\right)\\
W_{p}^-(u)=B_{p}^-D_{-i\gamma^-_p/\pi}\left(-\sqrt{2}\exp{\left(-i\pi/4\right)}u\right),
\end{eqnarray}
where $B_{p}^\pm$ are an arbitrary constants, which we choose so
that (20) and (21) fit the incident WKB wave outside the conversion
layer. As it is demonstrated in Appendix A $W_{p}^+$ and $W_{p}^-$
correspond to the incident ordinary and extraordinary waves,
respectively. The parameter $\gamma^\pm_p$ has meaning the length of
an evanescence layer (see the equation (A.3)). Its value depends on
the mode's number $\phi_p$ and combination of the parameter's $N$
sign and the direction of the process (see (28)). That is, first,
either of the mode tunnels through an evanescence layer with the
efficiency inherent itself, second, the conversion efficiency being
in 1D model the invariant under the conversion's direction reversal
and the poloidal position of the incident beam changing has no
longer the same property in 2D inhomogeneous plasma. Being mentioned
for the first time in~\cite{Ref8} for the oversimplified plasma
model ignoring the poloidal magnetic field and confirmed
in~\cite{Ref9} for the reasonable plasma model accounting for the
poloidal magnetic field $B_y/B\ll 1$ this property depends entirely
on the essential two-dimensional character of the waves.

Inserting (29) and (30) into (20), (21) and (11) after simple
algebra we obtain an integral representation of solution to (6)
\begin{eqnarray}
E_+({\bf r})=\exp{\left(in_z^0z+in_y^0y\right)}\sqrt{1+q_0}\times\nonumber\\
\int_{-\infty}^\infty{\frac{d\delta n_z}{2\pi}\exp{\left(i\delta
n_zz\right)}\left(\pm<I^\pm>+<R^\pm>\right)}\\
E_\xi({\bf r})=\exp{\left(in_z^0z+in_y^0y\right)}\sqrt{1+ia}\times\nonumber\\
\int_{-\infty}^\infty{\frac{d\delta n_z}{2\pi}\exp{\left(i\delta
n_zz\right)}\left(\mp<I^\pm>+<R^\pm>\right)}\nonumber\\
<...>=\int_{-\infty}^\infty
dy'{G\left(y-y'\right)\exp{\left(-i\tan{(\varphi)\delta n_z y'}
\right)}}\left(...\right),
\end{eqnarray}
where upper signs at $<I^\pm>$ correspond to the incident ordinary
wave and lower signs correspond to the incident extraordinary wave.
The explicit expressions for $I^\pm$ and $R^\pm$ are
\begin{eqnarray}
I^\pm=\sum_{k=0}^\infty{C_{k}\phi_{k}\left(\sqrt{|N|}v\right)D_{\pm i\gamma^\mp_k/\pi}\left(\pm\sqrt{2}\exp{\left(\pm i\pi/4\right)}u\right)},\,\,N>0\nonumber\\
I^\pm=\sum_{k=0}^\infty{C_{k}\phi_{k}\left(\sqrt{|N|}v\right)D_{\pm
i\gamma^\mp_k/\pi}\left(\pm\sqrt{2}\exp{\left(\pm
i\pi/4\right)}u\right)},\,\,N<0
\end{eqnarray}
and
\begin{eqnarray}
R^+=2|N|\exp{\left(
i\frac{\pi}{4}\right)}\times\nonumber\\
\sum_{k=0}^\infty {C_{k}k\phi_{k-1}(\sqrt{|N|}v)D_{
i\gamma^-_k/\pi-1}\left(\sqrt{2}\exp{\left(
i\pi/4\right)}u\right)},\,\,N>0\nonumber\\
R^+=-\sqrt{|N|}\exp{\left(
i\frac{\pi}{4}\right)}\times\nonumber\\
\sum_{k=0}^\infty {C_{k}\phi_{k+1}(\sqrt{|N|}v)D_{
i\gamma^-_k/\pi-1}\left(\sqrt{2}\exp{\left(
i\pi/4\right)}u\right)},\,\,N<0,\nonumber\\
R^-=2|N|\exp{\left(-
i\frac{\pi}{4}\right)}\times\nonumber\\
\sum_{k=0}^\infty {C_{k}k\phi_{k-1}(\sqrt{|N|}v)D_{-
i\gamma^+_k/\pi-1}\left(-\sqrt{2}\exp{\left(-
i\pi/4\right)}u\right)},\,\,N<0\nonumber\\
R^-=-\sqrt{|N|}\exp{\left(-
i\frac{\pi}{4}\right)}\times\nonumber\\
\sum_{k=0}^\infty {C_{k}\phi_{k+1}(\sqrt{|N|}v)D_{-
i\gamma^+_k/\pi-1}\left(-\sqrt{2}\exp{\left(-
i\pi/4\right)}u\right)},\,\,N>0.
\end{eqnarray}

While outside the mode conversion region at $u\rightarrow\pm\infty$
$<I^+>$ and $<I^->$ match to the geometrical optics solutions that
correspond to the WKB incident and conversed waves  (compare the
first terms in the r.h.s. of (A.1) and (A.2), (A.6) and (A.7)), the
WKB asymptotics of $<R^+>$ and $<R^->$ correspond to the WKB wave
reflected from the evanescence layer (see (A.5)).

It is possible to obtain asymptotic expansion of the solution for
$\varphi\rightarrow 0$. For the incident ordinary wave at $N>0$ we
have
\begin{eqnarray}
E_+({\bf r})=
\sqrt{1+q_0}\exp\left(in_z^0z\right)\int_{-\infty}^\infty\frac{d
\delta n_z}{2\pi}\sum_{k=0}^\infty\exp\left(i\delta
n_zz\right)C_{k}\times\nonumber\\(
\phi_{k}(\sqrt{|N|}v)D_{i\gamma^-_k/\pi}(\sqrt{2}\exp{(i\pi/4)}u)+\nonumber\\
2|N|\exp{\left(i\frac{\pi}{4}\right)} k\phi_{k-1}(\sqrt{|N|}v)D_{
i\gamma^-_k/\pi-1}(\sqrt{2}\exp{(
i\pi/4)}u))\\
E_\xi({\bf
r})=\sqrt{1+ia}\exp\left(in_z^0z\right)\int_{-\infty}^\infty\frac{d
\delta n_z}{2\pi}\sum_{k=0}^\infty\exp\left(i\delta
n_zz\right)C_{k}\times\nonumber\\(
-\phi_{k}(\sqrt{|N|}v)D_{i\gamma^-_k/\pi}(\sqrt{2}\exp{(i\pi/4)}u)+\nonumber\\
2|N|\exp{\left(i\frac{\pi}{4}\right)} k\phi_{k-1}(\sqrt{|N|}v)D_{
i\gamma^-_k/\pi-1}(\sqrt{2}\exp{( i\pi/4)}u)).
\end{eqnarray}
Last expressions recover the expressions for the electric field
components derived in~\cite{Ref9} for plasma with $B_y/B\ll 1$.

\section{The asymptotic expressions for the electric field's components}
For usual spherical tokamak's configuration O mode cut-off layer and
the X mode cut-off layer cross at the small angle, i.e. $a\ll 1$.
Keeping only terms in (23) being zero and first order quantity with
respect to the parameter $a$ we read it as
\begin{eqnarray}
x\simeq u-av/2,\nonumber\\
y\simeq v+au/2,
\end{eqnarray}
Then we simplify
\begin{eqnarray}
<I^\pm>\simeq\sum_{k=0}^\infty{...<\phi_{k}\left(y/d\right)}>,\nonumber\\
<R^\pm>\simeq\sum_{k=0}^\infty{...<\phi_{k\pm1}\left(y/d\right)}>,\nonumber
\end{eqnarray}
where
$d=\left(2^{1/4}(1+q_0)^{1/2}L_{by}^{1/2}\cos(\varphi)^{1/2}\right)/\left((2\pi\lambda)^{1/2}q_0^{1/4}\right)$
and $y$ is scaled in the units $\lambda=2\pi c/\omega$. Comparing
the asymptotic representation of $D_{\pm i\gamma_p/\pi}$ for the
argument $x\rightarrow\infty$ and asymptotic representation of
$D_{\pm i\gamma_p/\pi}$ for the argument $x\rightarrow -\infty$ (see
Appendix A) we obtain an explicit expressions for the conversion
coefficient, $T_{OX,XO}$, (in energy) and reflection coefficient,
$R_{OX,XO}$, where the subscribes $OX$ and $XO$ correspond to the
direction of the process
\begin{eqnarray}
T_{OX}=\frac{1}{P}\sum_{p=0}^\infty\int_{-\infty}^{\infty}\frac{d\delta
n_z}{2\pi}{|C_{p}|^{2}\exp{\left(-2\gamma_p^{\mp}\right)}},\,
R_{OX}=1-T_{OX},\\
P=\sum_{p=0}^\infty\int_{-\infty}^{\infty}\frac{d\delta
n_z}{2\pi}{|C_{p}|^{2}}\nonumber
\end{eqnarray}
and
\begin{eqnarray}
T_{XO}=\frac{1}{P}\sum_{p=0}^\infty\int_{-\infty}^{\infty}\frac{dn_z}{2\pi}{|C_{p}|^{2}\exp{\left(-2\gamma_p^{\pm}\right)}},\,
R_{XO}=1-T_{XO}.
\end{eqnarray}
The coefficient $C_n$ is given by the expression
\begin{eqnarray}
C_p=(2\pi)^2\times\nonumber\\
\int_{-\infty}^\infty{
A(y,z)\exp{\left(-i2\pi(n_z^0z+\delta
n_zz+n_y^0y)\right)}<\phi^*_p(y/d)>dydz},
\end{eqnarray}
where $*$ means the complex conjugation, $A(y,z)$ is given field
distribution in the incident beam in the WKB region, $x$ and $z$ are
scaled in $\lambda$ and $P$ is a normalizing coefficient
proportional to the beam intensity. The expressions (38) and (39)
derived similar to ones obtained in~\cite{Ref8, Ref9} excepting the
formula for $C_p$ given by (40) and deserve one comment. Noting
$T_{OX}\neq T_{XO}$ at the fixed $a$, we could read the reciprocal
relation  for the conversion coefficients in 2D inhomogeneous plasma
as
\begin{eqnarray}
T_{OX}\left({\bf B}\right)=T_{XO}\left(-{\bf B}\right),\,\,
R_{OX}\left({\bf B}\right)=R_{XO}\left(-{\bf B}\right),
\end{eqnarray}
where the argument ${\bf B}$ is presented explicitly. For the first
time the property (41) was mentioned in ~\cite{Ref8}. In
~\cite{Ref9} it was proved for plasma with poloidal magnetic field
$B_p/B\ll 1$.

For illustration, in figure 5 the conversion coefficient, $T_{OX}$,
of an incident beam with transversal distribution
$$A(y,z)=\left(\pi\rho^2\right)^{-1/4}\exp{\left(-y^2/(2\rho^2)+in^{opt}\sin{(\varphi)}y+in^{opt}\cos{(\varphi)}z\right)},$$
calculated with the use of 1D model~\cite{Ref4} (dashed curve) and
the formula (38) (solid curves) for $a<0$ and different $\varphi$
versus the dimensionless radius, $\rho_y/\lambda$, is shown.
Calculations are executed under conditions of EBW heating experiment
in MAST $L_n/\lambda=3$, $L_{by}/\lambda=70$, $q_0=0.7$,
$\omega/(2\pi)=15 GHz$. Rather strong variation of dependence
presented in figure 5 with variation of $\varphi$ is observed.

\section{Summary and conclusion}
The mode conversion of O mode to X mode have been examined in a
spherical tokamak geometry for plasma with a cold plasma dielectic
tensor. We have assumed the wavelengths are much shorter than the
equilibrium plasma gradient length, $L_n$. Thus, we expect WKB
approximation to apply except possibly near O-mode cut-off surface
or mode conversion layer.

Expanding wave equations in the region near the intersection of O
and X mode cut-off surfaces, one finds reduced wave equations
appropriate for this region which depends on two coordinates, $x$
and the poloidal angle $y$. We seek the required solution to the set
of equations so that it matches to the WKB solution outside the mode
conversion region. After functional substitution we have found the
integral representations of the required solution.

To the case typical of a spherical tokamak's configuration for which
the O mode cut-off layer and the X mode cut-off layer cross at the
small angle we have been simplified the obtained integral
representations of solution and found the wave fields and the
conversion coefficients explicitly.

We have shown that at the fixed poloidal position, i.e. fixed $a$,
the conversion coefficients obey the reciprocial relation (41) as it
was shown earlier in the frame of the model neglecting the poloidal
magnetic field ~\cite{Ref8} and of the model in which the poloidal
magnetic field is assumed to be small, $B_y/B\ll 1$.

The importance of 2D effects has been demonstrated by the example
under the MAST conditions.

\section{Acknowledgments} The work was supported by RFBR grants
04-02-16404, 06-02-17212

\begin{figure}
\includegraphics[height=60mm,bb= -150 0 310 240,clip]{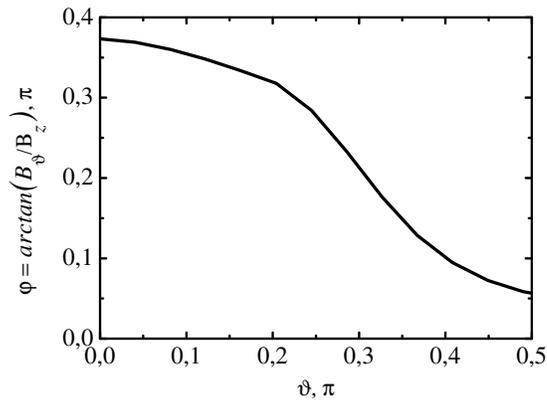}
\caption{$\varphi=\arctan{B_\theta/B_z}$ behavior versus the
poloidal angle $\vartheta$ for a typical MAST tokamak discharge}
\end{figure}
\begin{figure}
\includegraphics[height=60mm,bb= -150 356 390 630,clip]{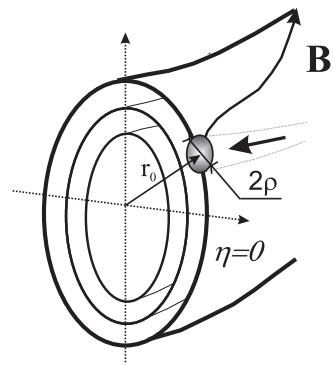}
\caption{Beam with radius $\rho$ of ordinary polarized
electromagnetic waves incident on the O-mode cut-off surface. The
co-ordinate ${\bf r}_0$ indicates the position of the beam center on
the O-mode cut-off surface}
\end{figure}

\begin{figure}
\includegraphics[height=60mm,bb= -125 485 540 800,clip]{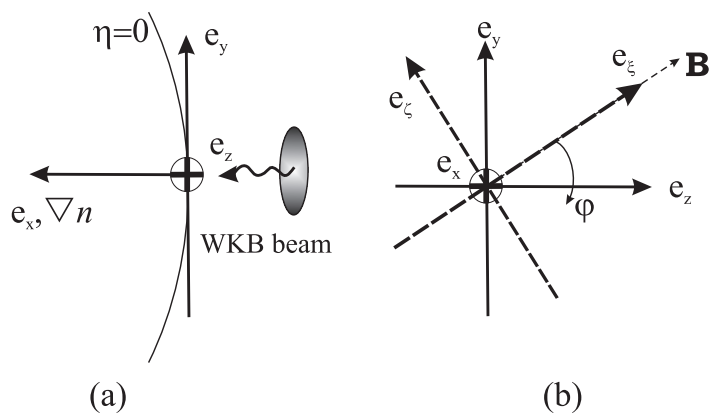}
\caption{Illustration of the co-ordinate systems
$\left(x,y,z\right)$ and $\left(x,\zeta,\xi\right)$ used. (a)
Two-dimensional subspace ($xoy$); (b) Two-dimensional subspace
($yoz$) with $\varphi$ being an angle between the magnetic field and
the toroidal direction}
\end{figure}

\begin{figure}
\includegraphics[height=60mm,bb=-100 480 590 745,clip]{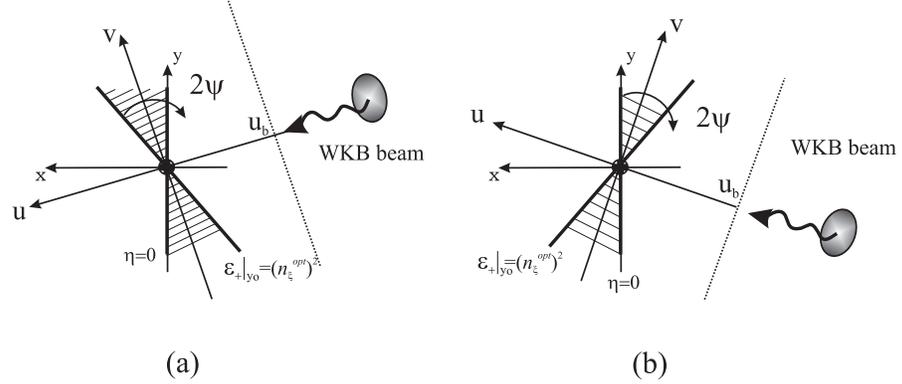}
\caption{Co-ordinate systems $\left(x,y\right)$ and
$\left(u,v\right)$ (a) $a>0$ ($N<0$); (b) (a) $a<0$ ($N>0$)}
\end{figure}

\begin{figure}
\includegraphics[height=60mm,bb= -40 126 147 226,clip]{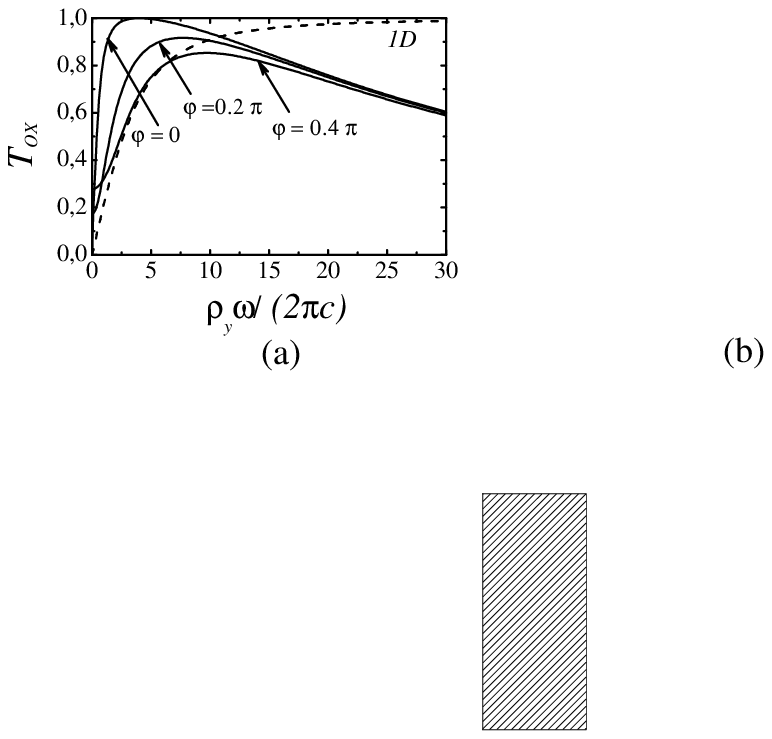}
\caption{The conversion coefficient, $T_{OX}$, versus the
dimensionless radius, $\rho_y/\lambda$, of an incident beam
calculated with the use of 1D model~\cite{Ref4} (dashed curve) and
the formula (38) (solid curves) for $a<0$ and different $\varphi$
under conditions of EBW heating experiment in MAST $L_n/\lambda=3$,
$L_{by}/\lambda=70$, $q_0=0.7$, $\omega/(2\pi)=15 GHz$.}
\end{figure}

\appendix

\section{Asymptotic expressions for the parabolic cylinder functions}
We pay our attention to the functions of the parabolic cylinder with
different arguments and indices
$D_{i\gamma/\pi}\left(\sqrt{2}\exp{\left(i\pi/4\right)}u\right)$ and
$D_{-i\gamma/\pi}\left(-\sqrt{2}\exp{\left(-i\pi/4\right)}u\right)$.
Let us begin from asymptotic expressions for the first one at two
limits $u\rightarrow\pm\infty$ which describe the corresponding WKB
solutions beyond the conversion layer. When the argument $u$ tends
to $+\infty$~\cite{Ref12}
\begin{eqnarray}
D_{i\gamma/\pi}\left(\sqrt{2}\exp{\left(i\pi/4\right)}u\right)|_{u\rightarrow\infty}\approx\left(\sqrt{2}u\right)^{i\gamma/\pi}
\exp{\left(-\gamma/4-iu^2/2\right)},
\end{eqnarray}
while at $u\rightarrow -\infty$
\begin{eqnarray}
D_{i\gamma/\pi}\left(\sqrt{2}\exp{\left(i\pi/4\right)}u\right)|_{u\rightarrow
-\infty}\approx\left(\sqrt{2}u\right)^{i\gamma/\pi}
\exp{\left(3\gamma/4-iu^2/2\right)}-\nonumber\\
\frac{\sqrt{2\pi}}{\Gamma\left(-i\gamma/\pi\right)}\frac{\exp{\left(i3\pi/4\right)}}{\left(\sqrt{2}u\right)^{1+i\gamma/\pi}}\exp{\left(\gamma/4+iu^2/2\right)}.
\end{eqnarray}
The second term in (A.2) is much smaller (with factor $1/u$)
compared to the first one. Therefore the function
$D_{i\gamma/\pi}\left(\sqrt{2}\exp{\left(i\pi/4\right)}u\right)$ in
the WKB sense describes the wave approaching an evanescence layer
from $-\infty$ and conversed wave propagating from an evanescence
layer to $\infty$. Comparing the coefficients at
$\exp{\left(-iu^2/2\right)}$ in (A.1) and (A.2) one obtains the
conversion coefficient (in energy) for the wave described by the
parabolic cylinder equation
\begin{eqnarray}
T=\exp{\left(-2\gamma\right)}.
\end{eqnarray}
Farther we obtain the derivative of the parabolic cylinder function
\begin{eqnarray}
\frac{\partial}{\partial
u}D_{i\gamma/\pi}\left(\sqrt{2}\exp{\left(i\pi/4\right)u}\right)=\nonumber\\
\frac{2i\gamma}{\pi}\exp{\left(i\pi/4\right)}D_{i\gamma/\pi-1}\left(\sqrt{2}\exp{\left(i\pi/4\right)}u\right)-iuD_{i\gamma/\pi}\left(\sqrt{2}\exp{\left(i\pi/4\right)}u\right)
\end{eqnarray}
The first term in the r.h.s. of (A.4), which has the asymptotics
~\cite{Ref12}
\begin{eqnarray}
D_{i\gamma/\pi-1}\left(\sqrt{2}\exp{\left(i\pi/4\right)}u\right)|_{u\rightarrow\infty}\approx
O(1/|u|)\nonumber\\
D_{i\gamma/\pi-1}\left(\sqrt{2}\exp{\left(i\pi/4\right)}u\right)|_{u\rightarrow
-\infty}\approx\nonumber\\
\frac{\sqrt{2\pi}}{\Gamma\left(1-i\frac{\gamma}{\pi}\right)}\frac{1}{\left(\sqrt{2}u\right)^{i\gamma/\pi}}\exp{\left(\gamma/4+iu^2/2\right)}+O(1/|u|),
\end{eqnarray}
describes the reflected wave.

Comparing the asymptotic expansions for the parabolic cylinder
function
$D_{-i\gamma/\pi}\left(-\sqrt{2}\exp{\left(-i\pi/4\right)}u\right)$
~\cite{Ref12}
\begin{eqnarray}
D_{-i\gamma/\pi}\left(-\sqrt{2}\exp{\left(-i\pi/4\right)}u\right)|_{u\rightarrow
\infty}\approx
\left(\sqrt{2}u\right)^{-i\gamma/\pi}\exp{\left(3\gamma/4\right)\exp\left(iu^2/2\right)}
-\nonumber\\
\frac{\sqrt{2\pi}}{\Gamma\left(i\gamma/\pi\right)}
\frac{\exp\left(-i3\pi/4\right)}{\left(\sqrt{2}u\right)^{1-i\gamma/\pi}}
\exp{\left(\gamma/4\right)}\exp{\left(-iu^2/2\right)}.
\end{eqnarray}
and
\begin{eqnarray}
D_{-i\gamma/\pi}\left(-\sqrt{2}\exp{\left(-i\pi/4\right)}u\right)|_{u\rightarrow
-\infty}\approx\nonumber\\
\left(\sqrt{2}u\right)^{-i\gamma/\pi}\exp{\left(-\gamma/4\right)\exp\left(iu^2/2\right)},
\end{eqnarray}
we note that it describes the wave which incident on an evanescent
layer from $u\rightarrow \infty$ (see the first term in (A.5)) and
propagates from an evanescent layer to the plasma boundary
($u\rightarrow -\infty$) (see (A.6)).

\end{document}